\documentclass[aps, prl, twocolumn, showpacs, floatfix,10pt,superscriptaddress]{revtex4-2}
\usepackage{amsmath}
\usepackage{amsfonts}
\usepackage{amsbsy}
\usepackage{amssymb}
\usepackage{graphicx}
\usepackage{textcomp}
\usepackage[caption=false,singlelinecheck=false]{subfig}
\usepackage{xcolor}
\usepackage{mathrsfs}
\usepackage{mathtools}
\usepackage{bm}
\usepackage{gensymb}
\usepackage{braket,wasysym}
\usepackage{float}
\usepackage{tikz}
\usepackage[colorlinks,linkcolor=blue,citecolor=red,filecolor=magenta,urlcolor=red,breaklinks]{hyperref}
\usepackage[bottom]{footmisc}
\usepackage{verbatim}

\hypersetup{colorlinks=true, urlcolor=blue, citecolor=red, pdfborder={0 0 0}}
\usepackage{breakurl}
\usepackage{natbib}

\allowdisplaybreaks

\newcommand{\tcb}[1]{\textcolor{blue}{#1}}

\normalfont
\def\be{\begin{equation}}
	\def\ee{\end{equation}}
\def\bea{\begin{eqnarray}}
	\def\eea{\end{eqnarray}}

\begin{document}
\title{
 Electron Model on Truchet Tiling:\\ Extended-to-Localization Transitions and Asymmetric Spectrum
 %Electron Model on Truchet Tiling:\\Extended-to-Localized Transitions, Mobility Edge, and Asymmetric Spectrum
}

\author{Junmo Jeon}
\email{junmojeon@sophia.ac.jp}
\affiliation{Department of Physics, Korea Advanced Institute of Science and Technology, Daejeon, 34141, Korea}
\affiliation{Physics Division, Sophia University, Chiyoda-ku, Tokyo 102-8554, Japan}
\author{Shiro Sakai}
\email{shirosakai@sophia.ac.jp}
\affiliation{Physics Division, Sophia University, Chiyoda-ku, Tokyo 102-8554, Japan}

\date{\today}
\begin{abstract}
Motivated by recent advances in the realization of Truchet-tiling structures in molecular networks and metal-organic frameworks, we investigate the wave localization issue in this kind of structure. We introduce an electron model based on random Truchet tilings—square lattices with randomly oriented diagonal links—and uncover a rich interplay between spectral and localization phenomena. By varying the strength of diagonal couplings, we explore successive transitions from an extended phase, through a regime with a mobility edge, to a fully localized phase. The energy-resolved fractal dimension analysis captures the emergence and disappearance of mobility edges, while an anomalous shift and asymmetry in the van Hove singularity are identified as key signatures of the underlying disordered Truchet-tiling structure. Notably, by using the finite-size scaling of level spacing statistics, we clarify that the transition occurs at a finite level of disorder even in the two-dimensional system. Our findings position Truchet-tiled electron systems as a versatile platform for engineering disorder-driven localization and interaction effects in amorphous quantum materials and photonic architectures.
\end{abstract}
\maketitle

\begin{figure}[h]
    \centering
    \includegraphics[width=0.45\textwidth]{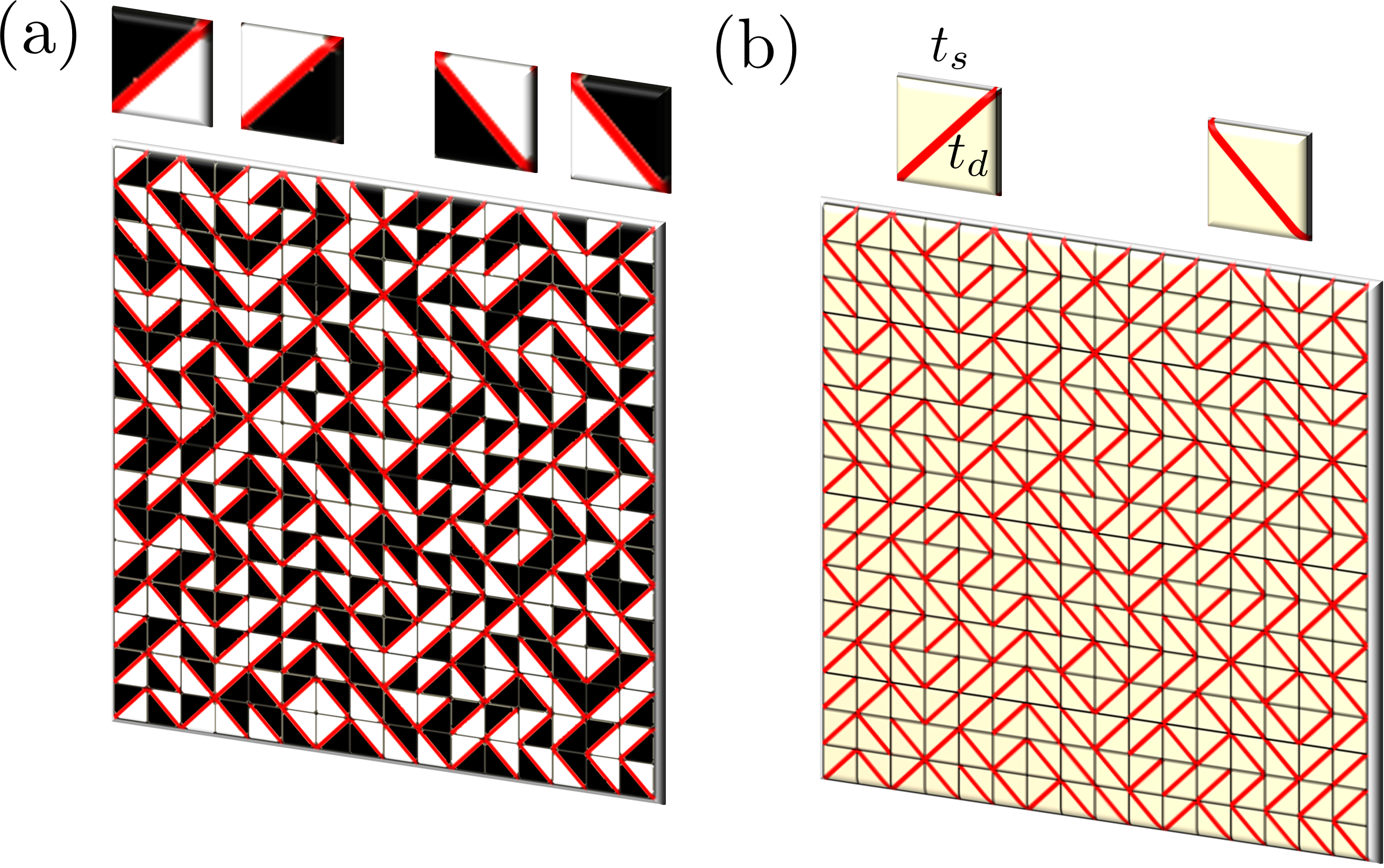}
    \caption{(a) Random Truchet-tiling composed of square tiles split diagonally into black and white triangles, each allowing four distinct orientations. (b) The corresponding tight-binding model on the square lattice, where two types of red diagonal links are introduced depending on the direction of diagonal split of each tile in (a). The black and red links represent two different hopping amplitudes, $t_s$ and $t_d$, respectively.}
    \label{fig: Truchet}
\end{figure}

\textit{\tcb{Introduction---}}
Truchet tiles are square tiles decorated with a rotationally asymmetric motif [see Fig.~\ref{fig: Truchet}(a)]\cite{truchet1704memoire}. When a plane is covered by these tiles with different orientations, various eye-catching patterns are generated.
Although this tiling had long been a purely theoretical construct with no counterpart in nature, it was recently realized in 
DNA molecular networks \cite{tikhomirov2017programmable} and metal-organic frameworks\cite{meekel2023truchet}.
These experimental realizations have
%The recent experimental realization of Truchet tiling structures in metal-organic framework\cite{meekel2023truchet} has 
sparked growing interest in engineered forms of disorder that can be precisely controlled and dynamically tuned\cite{wang2023synthesis,kenzhebayeva2024light,shi2024quasicrystal,li2025cocrystals}. They can form aperiodic and labyrinthine patterns, characterized by randomly oriented but structurally simple motifs\cite{meekel2023truchet,wang2023synthesis,yaghi2023decoding,smith1987tiling,adler2013physical}. Because of these characteristics, the Truchet-tiled structure—despite being aperiodic—can exhibit a range of distinctive properties that set it apart from conventional amorphous systems\cite{adler2013physical,griffin2025lanthanide,wang2023synthesis,jiang2024multivariate}. In particular, they enable systematic exploration of wave localization and spectral properties in aperiodic yet reproducible geometries\cite{wang2023synthesis,kenzhebayeva2024light,shi2024quasicrystal,li2025cocrystals}.

Wave localization in the absence of periodicity is a central topic in condensed matter and photonic physics, underpinning key phenomena such as Anderson localization and metal-insulator transitions\cite{anderson1958absence,abrahams1979scaling,segev2013anderson,lee1981anderson,imada1998metal,mott2004metal}. In Anderson localization, aperiodicity stems from random on-site energies within a bounded interval \cite{anderson1958absence}, while in a random Truchet tiling, it is introduced via the random placement of tiles selected from a finite set of orientations \cite{meekel2023truchet,smith1987tiling}.
%\tcr{***I do not understand the above sentence: While one can control the pattern of Truchet tiling, 'random' Truchet tiling is not controllable in the same sense as the random potential.***}
Thus, unlike conventional Anderson-type models, random Truchet tilings generate disordered networks whose physical properties can be analyzed using graph-theoretic tools\cite{blote1994fully}. Although structurally disordered systems of this type have been studied in statistical contexts such as percolation and loop models \cite{temperley2004relations, blote1994fully, nahum20113d, fendley2006loop}, their implications for wave localization and spectral properties remain largely unexplored.

In this paper, we introduce a simple electron model which captures the essence of the Truchet-tiling randomness, studying
the wavefunction localization and anomalous spectral properties in the model.
%a model based on random Truchet tilings: 
We consider square lattices decorated with randomly oriented diagonal hopping links of uniform strength [Fig.~\ref{fig: Truchet}(b)]. By tuning the strength of diagonal couplings, we explore the interplay between spectral and localization properties. We identify three distinct 
%disorder 
regimes: an extended phase, a phase with mobility edges, and a fully localized phase. The energy-resolved fractal dimension %framework 
captures the emergence and disappearance of mobility edges. Notably, finite-size scaling of level spacing statistics reveals that a localization transition occurs at a finite diagonal hopping strength.
%in the random Truchet tiling. 
This reflects the nature of two-dimensional Anderson localization as a marginal critical dimension, where extremely long localization lengths emerge under weak disorder\cite{abrahams1979scaling}.  Furthermore, we uncover anomalous shift and asymmetry in the van Hove singularity (vHs), providing insights into the connection between spectral features and localization phenomena. 
Our results position Truchet tilings as a promising platform for engineering disorder-driven effects in quantum materials.

\textit{\tcb{Random Truchet tiling---}} Truchet tiling is comprised of unit tiles that are 
%divided into two or more regions with simple shapes such as curves or diagonal splits
decorated with a motif breaking the rotational symmetry of the lattice
\cite{truchet1704memoire,meekel2023truchet,browne2008truchet,smith1987tiling}. These tiles are placed in such a way that they form visually complex patterns\cite{hall2019exploring,trevino2023aperiodic}. Figure \ref{fig: Truchet}(a) exhibits a random Truchet tiling composed of square tiles diagonally divided into black and white triangles, each randomly oriented among four possible orientations. %Note that the Truchet tiling in Fig.~\ref{fig: Truchet} (a), while lacking long-range order and being aperiodic, is comprised of four simple deterministic local tiles.
%\tcr{***I am not sure if 'deterministic local' is necessary.***}

Let us introduce the diagonal coupling along the boundary of black and white triangles in each plaquette [see Fig.~\ref{fig: Truchet}(b)]. 
The 
%Truchet 
tiling shown in Fig.~\ref{fig: Truchet}(b) is comprised of only two kinds of mutually exclusive decorated plaquettes: type-1 ($\tikz[baseline=-0.5ex]{
    \draw (0,0) rectangle (0.3,0.3);
    \draw[red, line width=0.5pt] (0.3,0) -- (0,0.3);
}$) and type-2 ($\tikz[baseline=-0.5ex]{
    \draw (0,0) rectangle (0.3,0.3);
    \draw[red, line width=0.5pt] (0,0) -- (0.3,0.3);
}$). 
%\tcr{The aperiodicity arises from the absence of long-range correlations between different plaquettes.[Even if it is quasiperiodic (long-range ordered), it is aperiodic... Do we need this and next sentences?]} 
%We emphasize that this structural feature serves as a key distinction between the random Truchet-tiling pattern and conventional amorphous models with uniformly random potentials or hopping amplitudes, as well as quasiperiodic systems possessing long-range order\cite{stachurski2011structure,hiramoto1989new,jagannathan2021fibonacci}.

Such exotic structure of Truchet tiling could be realized experimentally, for example, via metal-organic framework\cite{meekel2023truchet}, DNA molecular networks\cite{tikhomirov2017programmable}, ultracold atoms\cite{julia2024amorphous}, patterned photonic crystals\cite{hajshahvaladi2022very,chen2012fabrication} or by engineered electronic circuits in which resonant elements are connected along randomized diagonals\cite{helbig2019band,kandangath2007inducing,todorov2002tight}. To probe the resulting electronic—or, more generally, wave—dynamics on this unconventional geometry, we consider a tight-binding model in which, in addition to conventional nearest-neighbor hoppings in square lattice, a diagonal hopping term is introduced between the two sites of each plaquette. For instance, in ultracold‐atom experiments, one can use site-selective Raman‐assisted tunneling to imprint a two‐photon resonance along one diagonal of each plaquette of square optical lattice, thereby giving each plaquette an independent next-nearest-neighbor hopping amplitude and realizing the Truchet-tiling tight-binding model\cite{miyake2013realizing,carpentier2010raman,aidelsburger2011experimental}.% In photonic platforms, silicon photonic-crystal slabs with SRRs on each square unit cell support localized modes, where the orientation and diagonal coupling of SRRs can be independently controlled, directly mimicking random diagonal hoppings in a Truchet tiling lattice\cite{bogdanov2019bound,bulu2005compact,segal2015controlling}.

The tight-binding Hamiltonian on the 
%Truchet-tiled system 
Truchet tiling
is given by
\begin{align}
\label{H}
&H=-t_s\sum_{\braket{i,j}\sigma}c_{i\sigma}^\dagger c_{j\sigma}-t_d\sum_{\braket{\braket{i,j}}_{\mathrm{diag}}\sigma}c_{i\sigma}^\dagger c_{j\sigma}+\mathrm{H.c.},
\end{align}
where $t_s$ and $t_d$ are the amplitudes for horizontal/vertical and diagonal hopping amplitudes, respectively. $c_{i\sigma}$ and $c_{i\sigma}^\dagger$ are electronic annihilation and creation operator at the $i$ site and spin $\sigma$, respectively. $\braket{\braket{i,j}}_{\mathrm{diag}}$ runs over the randomly chosen diagonal bonds. We adopt open boundary conditions; however, our findings can also be observed under periodic boundary conditions\cite{supplementTruchet1}. The strength of disorder inherent in the diagonal pattern is given by the ratio $t_d/t_s$. To systematically explore the influence of unique geometry on the wave localization, we reparametrize the model by setting $t_d=\alpha t$ and $t_s=(1-\alpha)t$, with $0\le \alpha< 1$. We set $t=1$. By varying $\alpha$, the electron localization properties of full range of disorder strengths inherent in the Truchet-tiling pattern can be explored. We perform our calculations over 1000 independent realizations of the random Truchet tiling and average them.

\textit{\tcb{Anomalous van Hove shift---}} Before delving into the localization characteristics of the electron states on the Truchet tiling, we explore the notable features of the density of states (DoS), $\rho(E)=\sum_n\delta(E-E_n)$, where $E_n$ is the $n$-th eigenenergy. We focus on the DoS of a random Truchet tiling and compare it with that of an anisotropic triangular (AT) lattice, a long-range ordered counterpart.\cite{avgin2017absorption,suros2008study}. Note that the AT lattice is comprised of only one of two tile orientations in Fig.~\ref{fig: Truchet}(b).

\begin{figure}[h]
    \centering
    \includegraphics[width=0.4\textwidth]{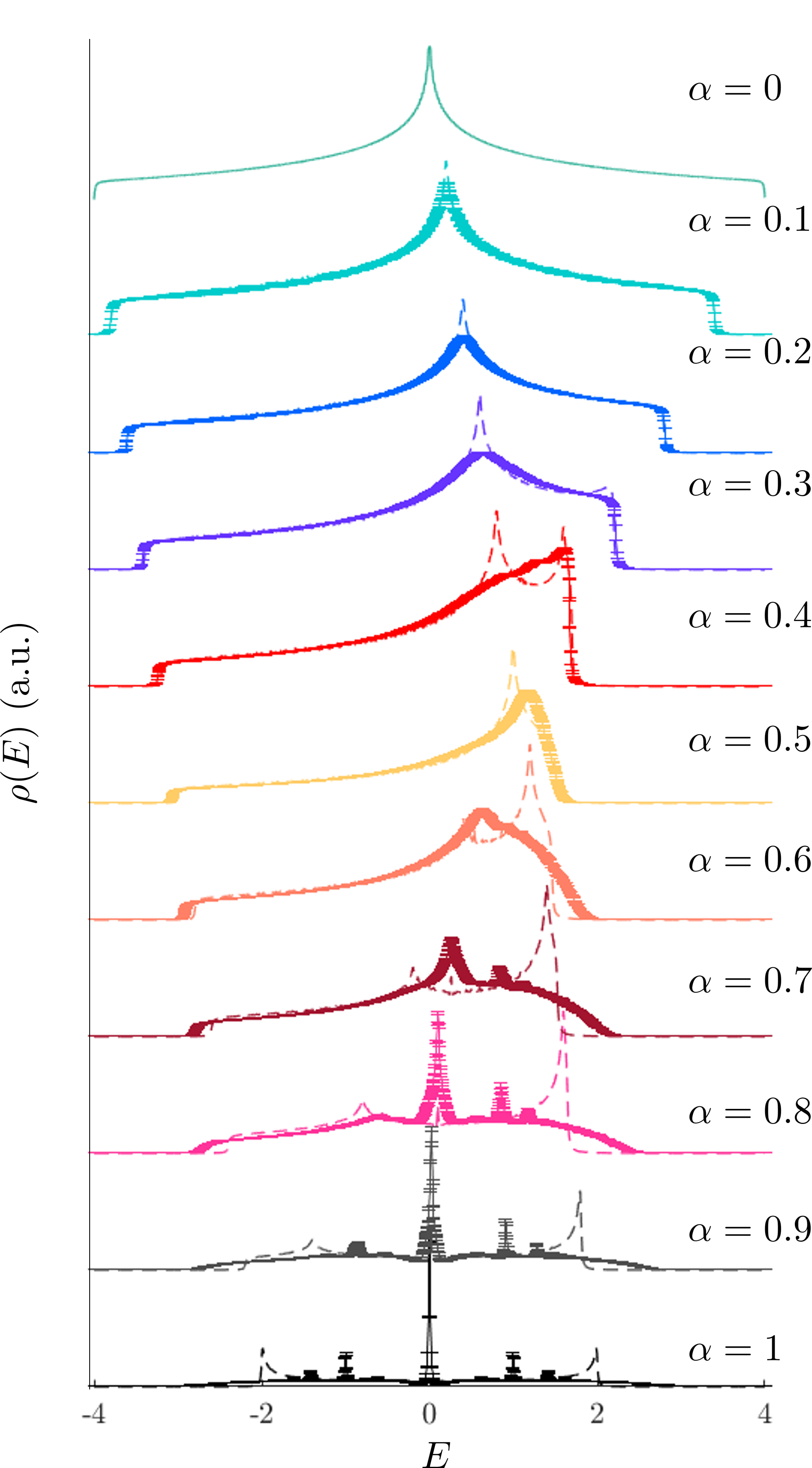}
    \caption{Density of states in arbitrary units of random Truchet tilings (solid curves with error bar) and anisotropic triangular lattice (dashed curves) for different $\alpha$ values. For $\alpha=0$, the system is simple square lattice. (Random Truchet tilings) As $\alpha$ increases, the energy at which the 
    %van Hove singularity 
    vHs
    appears initially starts at zero and gradually shifts upward, reaching the top of the energy band near $\alpha\approx 0.4$. Beyond this point, it decreases and approaches zero again. For $0.4\le\alpha<0.7$, the 
    %density of states 
    DoS
    in the vicinity of the 
    %van Hove singularity 
    vHs
    becomes markedly asymmetric. (Anisotropic triangular lattice) There are two 
    %van Hove singularities 
    vHs
    in general. For every $\alpha$, the vicinity of the 
    %van Hove singularity
    vHs
    is symmetric. As $\alpha$ increases, the %van Hove singularities 
    vHs
    monotonically shift in opposite energy directions to the spectral edges. The system size is $130\times130$.}
    \label{fig: dos}
\end{figure}

\begin{figure}[h]
    \centering
    \includegraphics[width=0.5\textwidth]{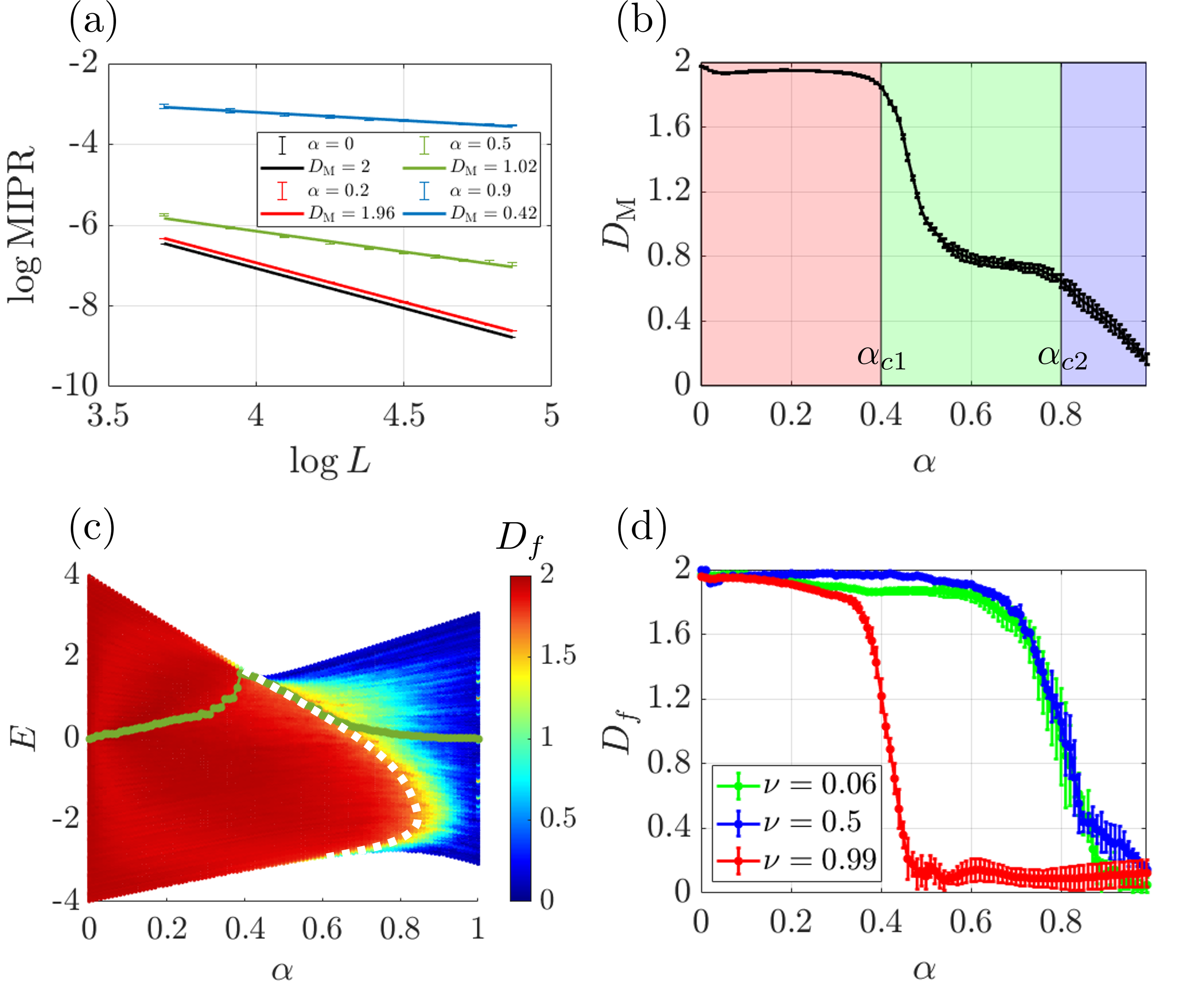}
    \caption{Transition of the electron localization property in the Truchet tiling. (a) Power-law scaling behavior of the mean IPR (MIPR) for different $\alpha$ values as the function of $L$. $40\le L\le 130$. (b) Three different electron %localization 
    phases in terms of the mean fractal dimension, $D_\mathrm{M}$ as %the
    a
    function of $\alpha$. In the red shaded region ($0\le \alpha\le \alpha_{c1}$), $D_\mathrm{M}$ stays close to two even for nonzero $\alpha$. In the green shaded region ($\alpha_{c1}<\alpha\le\alpha_{c2}$), $D_\mathrm{M}$ drops to 0.8 and remains nearly constant. $D_\mathrm{M}$ decreases again in the blue shaded region ($\alpha>\alpha_{c2}$). $\alpha_{c1}\approx0.4$ and $\alpha_{c2}\approx0.8$. (c) The landscape of the generalized fractal dimension ($D_f$) as %the 
    a
    function of energy and 
    %disorder parameter 
   $\alpha$. The white dashed and green solid curves are drawn 
    %for emphasizing 
   to emphasize
    the mobility edge and 
    %van Hove singularity
    vHs
    ($E_{\mathrm{vHs}}$), respectively. (d) The generalized fractal dimension at the Fermi energy for given filling fraction, $\nu$, as 
    %the 
    a
    function of $\alpha$. The critical value of $\alpha$, where the generalized fractal dimension rapidly drops, 
    becomes small 
    %when the filling fraction $\nu$ is close to unity. 
    for $\nu\simeq 1$.
    %Here, $40\le L\le 130$.
    }
    \label{fig: fractalD}
\end{figure}

Figure \ref{fig: dos} compares DoS of the random Truchet tiling and AT lattice for different $\alpha$ values. Remind that the square lattice ($\alpha=0$) has a symmetric log-divergent %van Hove singularity 
vHs
at the zero energy. First of all, we observe that the number of 
%van Hove singularities 
vHs
of the random Truchet tiling remains one for nonzero $\alpha$, unlike a general case of an AT lattice with two 
%van Hove singularities
vHs
%in general
\cite{suros2008study,cohen2016fundamentals}. Second, interestingly, the 
%van Hove singularity 
vHs
of the random Truchet tiling shows anomalous non-monotonic shift as a function of $\alpha$. In detail, the 
%van Hove singularity 
vHs
moves toward higher energy until $\alpha=0.4$, and %it 
shifts back toward zero energy as 
%$\alpha\gtrsim0.4$ increases. 
$\alpha$ increases further.
Notably, the van Hove shift exhibits a sudden change around $\alpha=0.4$, marking a turning point that drastically alters the behavior of the shift [see green curve in Fig.~\ref{fig: fractalD}(c)]. This %is contrary to
contrasts with
 the van Hove shifts in the AT lattice, where two 
%van Hove singularities 
vHs
always shift monotonically in opposite energy directions with increasing $\alpha$, and 
%they smoothly coincide 
merge
at $\alpha=0.5$\cite{supplementTruchet1}. 
Furthermore, while the 
%van Hove singularities 
vHs
of the AT lattice are symmetric in their vicinity for general $\alpha$, in the random Truchet tilings, the 
%van Hove singularity 
vHs
for $0.4\le\alpha<0.7$ exhibits an 
%asymmetry in its 
asymmetric
peak shape, in contrast to the cases of $\alpha<0.4$ and $\alpha\ge0.7$ (compare the blue, orange and brown curves in Fig.~\ref{fig: dos}, for instance). 
%We note that such asymmetry is reminiscent of that observed in face-centered cubic lattices\cite{cohen2016fundamentals,loly1972density}. 
Third, although both systems recover the %bipartite feature 
bipartiteness
as $\alpha$ approaches unity, in contrast to the AT lattice, where two 
%van Hove singularities 
vHs
appear at the spectral edges, the most prominent 
%van Hove singularity 
vHs
in the random Truchet tiling moves to zero energy.
%as $\alpha\to 1$. 

Interestingly, unlike the 
%van Hove singularity 
vHs
at the band edge of the AT lattice, the zero-energy 
%van Hove singularity 
vHs
in the random Truchet tiling is protected by the chiral symmetry restored as $\alpha\to1$. Note that this chiral symmetry stems from the %bipartite nature
bipartiteness
of the graph formed solely by diagonal links and appears in both the AT lattice and the random Truchet tiling. However, while the AT lattice reduces to multiple decoupled one-dimensional chains, the random Truchet tiling develops many disconnected diagonal-link graphs that individually possess features such as loop-induced symmetry within a sublattice or sublattice imbalance, thereby hosting a macroscopically large number of symmetry-protected zero-energy states\cite{suros2008study,lieb1989two,vidal1998aharonov,kremer2020square,supplementTruchet1}.

%The emergence of a large number of zero-energy states in the random Truchet tiling as $\alpha\to1$ stems from either sublattice imbalance or symmetry within a sublattice in graphs formed by diagonal links. %See Supplemental Materials for detailed information\cite{supplementTruchet1} (see also references \cite{lieb1989two,vidal1998aharonov,kremer2020square} therein).
%Note that in the thermodynamic limit, both origins appear numerous times in the random Truchet tiling because they are factorized into many disconnected bipartite graphs with local squares as $t_s$ becomes negligible (see Fig.~\ref{fig: Truchet} (b)), whereas the AT lattice reduces to infinitely many decoupled one-dimensional chains\cite{suros2008study,supplementTruchet1}.

We emphasize that, beyond a certain threshold of disorder in random Truchet tilings, the resulting asymmetry in both the 
%van Hove singularity 
vHs
and DoS, reminiscent of that observed in face-centered cubic lattices\cite{cohen2016fundamentals,loly1972density}, can lead to notable consequences in the presence of electron correlations. 
This kind of asymmetry can stabilize ferromagnetism by suppressing Fermi surface nesting and inhibiting antiferromagnetic ordering, both of which generically rely on a symmetric band structure and well-nested Fermi surfaces\cite{fernando1997stoner,stoeffler1991strong,kanamori2001general}. Moreover, the asymmetric 
%van Hove singularity 
vHs
near the Fermi level significantly enhances the DoS at the Fermi energy, $\rho(E_F)$, thereby reducing the critical interaction strength required for ferromagnetic instability according to the Stoner criterion, $U\rho(E_F)>1$, where $U$ is the interaction strength\cite{fernando1997stoner}. As a result, the system becomes more susceptible to ferromagnetic ordering.

\textit{\tcb{Wave localization and mobility edge---}} Now let us explore the localization characteristics of electron wavefunctions. To identify the degree of localization of the wavefunction, we adopt the inverse participation ratio (IPR). The IPR of the eigenstate $\ket{\psi_n}$ is given by $\mathrm{IPR}_n=\sum_i \vert \psi_n(i)\vert ^4$, where $\psi_n(i)$ is the normalized wavefunction of $n$-th eigenstate at site $i$\cite{wegner1980inverse}. For uniformly extended states, IPR$\sim N^{-1}$ for large system size $N=L^2$ in two-dimensional systems, while IPR of the localized states is almost independent of the entire system size\cite{wegner1980inverse}. Generally, IPR$_n\sim L^{-D_n}$, where $D_n$ is the fractal dimension\cite{fyodorov1993level,jagannathan2021fibonacci}. In two-dimensional system, $D_n=0$ (2) for localized (extended) states. When $0<D_n<2$, the state is called critical, neither fully extended nor exponentially localized\cite{jeon2023localization,bauer1990correlation,jagannathan2021fibonacci}.

We first consider the mean IPR (MIPR) given by $\mathrm{MIPR}=\frac{1}{N}\sum_{n=1}^{N}\mathrm{IPR}_n$ as the function of system size\cite{jeon2023localization}. Figure \ref{fig: fractalD}(a) shows that MIPR also follows similar scaling behavior, $\mathrm{MIPR}\sim L^{-D_\mathrm{M}}$. Here, $D_\mathrm{M}$ is the mean fractal dimension, which represents the average localization property of the system. Figure \ref{fig: fractalD}(b) illustrates the mean fractal dimension as %the 
a
function of $\alpha$. We figure out three different phases in terms of distinct behaviors of $D_\mathrm{M}$. First, $D_\mathrm{M}$ stays close to 2 up to 
%a certain $\alpha$ value, $\alpha_{c1}\approx 0.4$.
$\alpha=\alpha_{c1}\approx 0.4$.
This indicates that most states behave like conventional extended state below this critical disorder strength 
[red shaded region in Fig.~\ref{fig: fractalD}(b)]. Second, $D_M$ decreases to 0.8 beyond $\alpha_{c1}$ and remains nearly constant until the second critical value, $\alpha_{c2}\approx 0.8$. As we will show, the states in this phase exhibit energy-dependent localization properties with a nontrivial mobility edge. Thus, the localized, extended and critical states coexist in this phase (green shaded region).
%in Fig.~\ref{fig: fractalD}(b)).
Lastly, for $\alpha>\alpha_{c2}$, we observe that $D_\mathrm{M}$ decreases again as $\alpha$ increases (blue shaded region). 
%in Fig.~\ref{fig: fractalD}(b)].
It turns out that this second drop in $D_\mathrm{M}$ is attributed to the absence of extended states in the spectrum. In the thermodynamic limit, the localized states form an Anderson insulator with an everywhere dense spectrum\cite{supplementTruchet1}.

To unveil the exotic phase diagram 
%explored 
shown
in Fig.~\ref{fig: fractalD}(b), 
%let us 
we
explore the energy-dependent localization characteristics. We investigate the average IPR of states in the vicinity of a given energy $E$ as a function of system size. In detail, we introduce the density of IPR, say DoIPR at the given energy $E$ as
\begin{align}
    \label{DoIPR}
    &\mathrm{DoIPR}(E)=\frac{1}{\pi}\sum_n \mathrm{IPR}_n\frac{\Gamma}{(E-E_n)^2+\Gamma^2},
\end{align}
where $\Gamma$ is %half of 
half
the linewidth. Note that %the 
DoIPR is an averaged IPR of the states in the vicinity of a given energy $E$ by using the Lorentzian function. We set $\Gamma=10^{-2}t$. The DoIPR($E$) follows the scaling behavior of $L^{-D_f(E)}$ for large system sizes $N=L^2$. Here, $D_f(E)$ is the generalized fractal dimension at the energy $E$ given by
\begin{align}
    \label{FD}
    &D_f(E)=-\lim_{N\to\infty}\frac{\log \mathrm{DoIPR}(E)}{\log L}
\end{align}
When the states around energy $E$ are ideally localized (extended), $D_f(E)=0$ (2), respectively. While, $0<D_f(E)<2$ implies either the coexistence of extended and localized states near $E$, or the presence of sufficiently many critical states. We emphasize that the generalized fractal dimension can be used to capture the presence of mobility edge, since it generally depends on the energy $E$ as we will show.

Figure \ref{fig: fractalD}(c) exhibits generalized fractal dimensions as 
%the 
a
function of energy and disorder parameter, $\alpha$. Note that until the disorder parameter exceeds the first critical value, $\alpha_{c1}\approx0.4$, $D_f$ remains close to 2 for general energies. As $\alpha>\alpha_{c1}$, the localized states appear in the spectrum, and hence $D_\mathrm{M}$ significantly decreases at $\alpha\gtrsim \alpha_{c1}$. Nevertheless, we emphasize that even for $\alpha>\alpha_{c1}$, there are extended states and critical states that mitigate the sharp decline of $D_\mathrm{M}$ toward zero. Specifically, Fig.~\ref{fig: fractalD}(d) 
%shows the generalized fractal dimension 
plots $D_f$
at the given filling fraction, $\nu$, as 
%the 
a
function of disorder parameter $\alpha$. Notably, the high-energy states become localized even for smaller $\alpha$ values, while the low-energy states require larger values of $\alpha$ to be localized. Thus, the system possesses a nontrivial mobility edge drawn as the white dashed curve in Fig.~\ref{fig: fractalD} (c). We note that this mobility edge suddenly disappears at $\alpha\gtrsim\alpha_{c2}\approx0.8$. This implies that the absence of nearly extended states—previously acting as a buffer—leads to a more rapid decline in $D_\mathrm{M}$ for $\alpha>\alpha_{c2}$.

Let us point out a notable relationship between the mobility edge and the 
%van Hove singularity.
vHs.
The green solid curve in Fig.~\ref{fig: fractalD}(c) represents the energy, $E_{\mathrm{vHs}}$, at which the most prominent %van Hove singularity 
vHs
appears. $E_\mathrm{vHs}$ in random Truchet tiling increases with $\alpha$ until $\alpha_{c1}$, then decreases, with a divergent slope at $\alpha_{ 
c1}$. Importantly, the 
%van Hove singularity 
vHs
exhibits a pronounced asymmetry in the DoS for $\alpha_{c1}\le\alpha<0.7$ (see Fig.~\ref{fig: dos}). This asymmetry in the DoS vicinity of the 
%van Hove singularity 
vHs
originates from the reorganization of localized and delocalized states that admit different level spacing statistics and the shape of DoS\cite{argyrakis1992density,mondal2019optical,devakul2017anderson}. Hence, the mobility edge, separating the nearly extended and localized states, is found to coincide with the energy of the asymmetric 
%van Hove singularity 
vHs
that emerges
for $\alpha_{c1}\le\alpha<0.7$.

    We emphasize that the anomalous extended-to-localized transition in the random Truchet tiling originates from structural constraints that have no counterpart in conventional random models. In typical disordered systems, on-site potentials or hopping terms are assigned independently, producing uncorrelated local environments. In contrast, the Truchet tiling imposes a plaquette-level rule: only one of two diagonal links can be present in each plaquette. This restriction limits local coordination numbers and suppresses large-scale structural fluctuations, resulting in a hyperuniform structure\cite{torquato2018hyperuniform,inpreparation}, which in turn delays the onset of Anderson localization. Consequently, two-dimensional extended states persist far beyond the range expected for uncorrelated disorder, disappearing only when the square-lattice hopping $t_s$ is sufficiently reduced\cite{supplementTruchet1}.% These findings uncover a previously unexplored mechanism—structurally constrained randomness—for stabilizing delocalized phases in disordered two-dimensional systems.

\textit{\tcb{Level spacing statistics---}}
The localization transition and spectral properties are in general deeply intertwined. To clarify their relationship in our model, we analyze the level spacing statistics using the average adjacent gap ratio, $\langle r\rangle$. The adjacent gap ratio is defined as $r_n = \min(\delta_n, \delta_{n-1}) / \max(\delta_n, \delta_{n-1})$, where $\delta_n = E_{n+1} - E_n$. The average $\langle r\rangle$ is then obtained by averaging $r_n$ over the spectrum. This provides a sensitive measure of level correlations\cite{devakul2017anderson,ndawana2002finite}. Localized phases without level repulsion follow Poisson statistics with $\langle r \rangle\approx 0.386$, while delocalized phases exhibit Wigner-Dyson (GOE) statistics with $\langle r \rangle\approx 0.531$ due to the level repulsion\cite{devakul2017anderson}. Assuming simple scaling form, $\langle r\rangle (\alpha, L)=f((\alpha-\alpha_T)L^{1/\mu})$ for some universal function $f$, $\langle r\rangle$ curves for different $L$ crosses at $\alpha=\alpha_T$. Here, $\mu$ and $\alpha_T$ are the critical exponent and scale-invariant transition point, respectively.

\begin{figure}[h]
    \centering
    \includegraphics[width=0.5\textwidth]{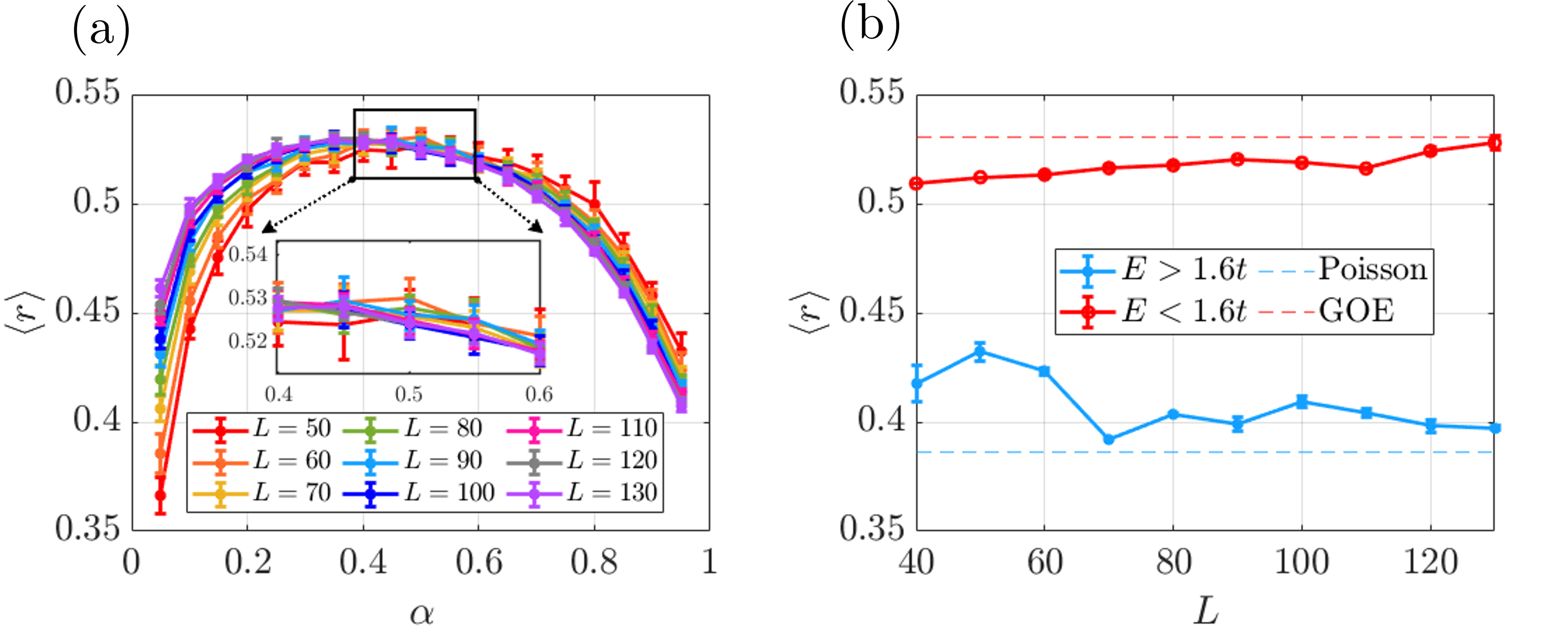}
    \caption{Level spacing statistics in the random Truchet tiling. (a) Finite-size scaling of the average adjacent gap ratio, $\langle r\rangle$, as a function of $\alpha$. As $L$ increases, $\langle r\rangle$ increases (decreases) for $\alpha<0.4$ ($\alpha>0.6$). The inset highlights the crossing of curves for different system sizes at $0.4\le\alpha_T\le0.6$. (b) $\langle r\rangle$ for two different energy windows as 
    %the
    a function of system size, calculated at
    %Here,
    $\alpha=0.7$. The red and blue curves represent $\langle r\rangle$ for different energy windows, $E<1.6t$ and $E>1.6t$, respectively. The red and blue dashed lines are drawn to emphasize the reference values of the Wigner-Dyson (GOE) and Poisson statistics, respectively.
    }
    \label{fig: FSS}
\end{figure}
Figure \ref{fig: FSS}(a) exhibits the average adjacent gap ratio for different system sizes. $\langle r \rangle$ increases (decreases) as the system size increases for $\alpha<0.4$ ($\alpha>0.6$), approaching the characteristic value of the Wigner-Dyson (Poisson) statistics. Thus, $\langle r \rangle$ curves for different system sizes crosses at $0.4\le \alpha_T\le 0.6$ (see the inset of Fig.~\ref{fig: FSS}). Such a crossing is reflected in Fig.~\ref{fig: fractalD}(b) as a sharp drop in the $D_M$ for $0.4\le \alpha\le 0.6$. Moreover, $\langle r\rangle$ approaches distinct characteristic values with increasing system size, depending on the energy window [see Fig.~\ref{fig: FSS}(b)]. These results highlight the interplay between localization and spectral properties.

\textit{\tcb{Conclusion---}}
In this work, motivated by the recent experimental realization of exotic Truchet-tiling structure, we unveiled a rich landscape of spectral and localization phenomena in an electron model based on random Truchet tilings—square lattices with randomly oriented diagonal hoppings. By varying the strength of diagonal %couplings, 
hoppings,
we clarified the transitions from an extended phase, through a regime with mobility edges, to an entirely localized phase in two dimensions. The energy-resolved fractal dimension analysis captured the emergence and disappearance of mobility edges, revealing a direct link between spectral asymmetry, mobility-edge dynamics, and van Hove singularities. Furthermore, the finite-size scaling of the level spacing statistics unveiled the intertwined spectral property and localization transitions. Our findings establish Truchet-tiling electron systems as a versatile framework for engineering disorder-driven localization and interaction effects in amorphous quantum materials and photonic platforms.

Looking ahead, our results open promising avenues for experimental applications in various platforms including metal-organic framework, ultracold atomic lattices and photonic crystals where dynamically programmable disorder can be used to control transport, coherence, and correlations\cite{kshetrimayum2021quantum,smith2016many,dai2024programmable}. Extending Truchet tilings to include interactions could enable novel phases and disorder-driven quantum technologies.

\section*{Acknowledgments}
We thank Tomi Ohtsuki for useful comments and valuable discussions. This work was supported by JSPS KAKENHI Grant No. JP25H01397 and 25H01398. J.M.J was supported by National Research Foundation Grant (2021R1A2C109306013) and Nano Material Technology Development Program through the National Research Foundation of Korea(NRF) funded by Ministry of Science and ICT (RS-2023-00281839).

\bibliography{reference}

\newpage
\begin{widetext}
\title{Supplemental Material 
%of
for
`Electron Model on Truchet Tiling:\\Extended-to-localized transitions and asymmetric spectrum'}

\author{Junmo Jeon}
\email{junmojeon@sophia.ac.jp}
\affiliation{Korea Advanced Institute of Science and  Technology, Daejeon 34141, South Korea}
\affiliation{Physics Division, Sophia University, Chiyoda-ku, Tokyo 102-8554, Japan}
\author{Shiro Sakai}
\email{shirosakai@sophia.ac.jp}
\affiliation{Physics Division, Sophia University, Chiyoda-ku, Tokyo 102-8554, Japan}

\maketitle

\section{Van Hove shifts in the anisotropic triangular lattice}
    \label{sec:1}
    We investigate the Van Hove singularity that arises in the anisotropic triangular (AT) lattice, a periodic Truchet tiled structure consisting solely of one type of diagonal link (see Fig.~\ref{fig: sup_AT} (a)). Since the system is periodic, the bulk Van Hove singularity can be identified by using the periodic boundary condition. When the horizontal/vertical and diagonal hopping amplitudes are $1-\alpha$ and $\alpha$, respectively, the energy dispersion relation is given by
    \begin{align}
        \label{dispersion_AT}
        &E(k_x,k_y)=-2(1-\alpha)(\cos k_xa+\cos k_ya)-2\alpha(\cos(k_x+k_y)a).
    \end{align}
    Here, $a$ is the lattice constant of the square lattice. Note that for $\alpha=0$ (1), the system becomes a square lattice and a one-dimensional chain, respectively. Then, the Van Hove singularities appear at $(k_xa,k_ya)=(\pi,0),(0,\pi)$ and $(\pi,\pi)$. Since $(k_xa,k_ya)=(\pi,0)$ and $(0,\pi)$ are degenerated, there are two Van Hove singularities at the energies, $E_{\mathrm{vHs}}=2\alpha$ and $4-6\alpha$. Thus, two Van Hove singularities shift monotonically in opposite energy directions with increasing $\alpha$, and they smoothly coincide at $\alpha=0.5$. This is contrary to the anomalous Van Hove shift in the random Truchet tiling, where the characteristics of the Van Hove shift drastically change
    and $dE_\mathrm{vHs}/d\alpha$ diverges at $\alpha_{c1}\approx0.4$ (compare Figs.~\ref{fig: sup_AT} (b) and (c)).

    \begin{figure}[h]
    \centering
    \includegraphics[width=0.6\textwidth]{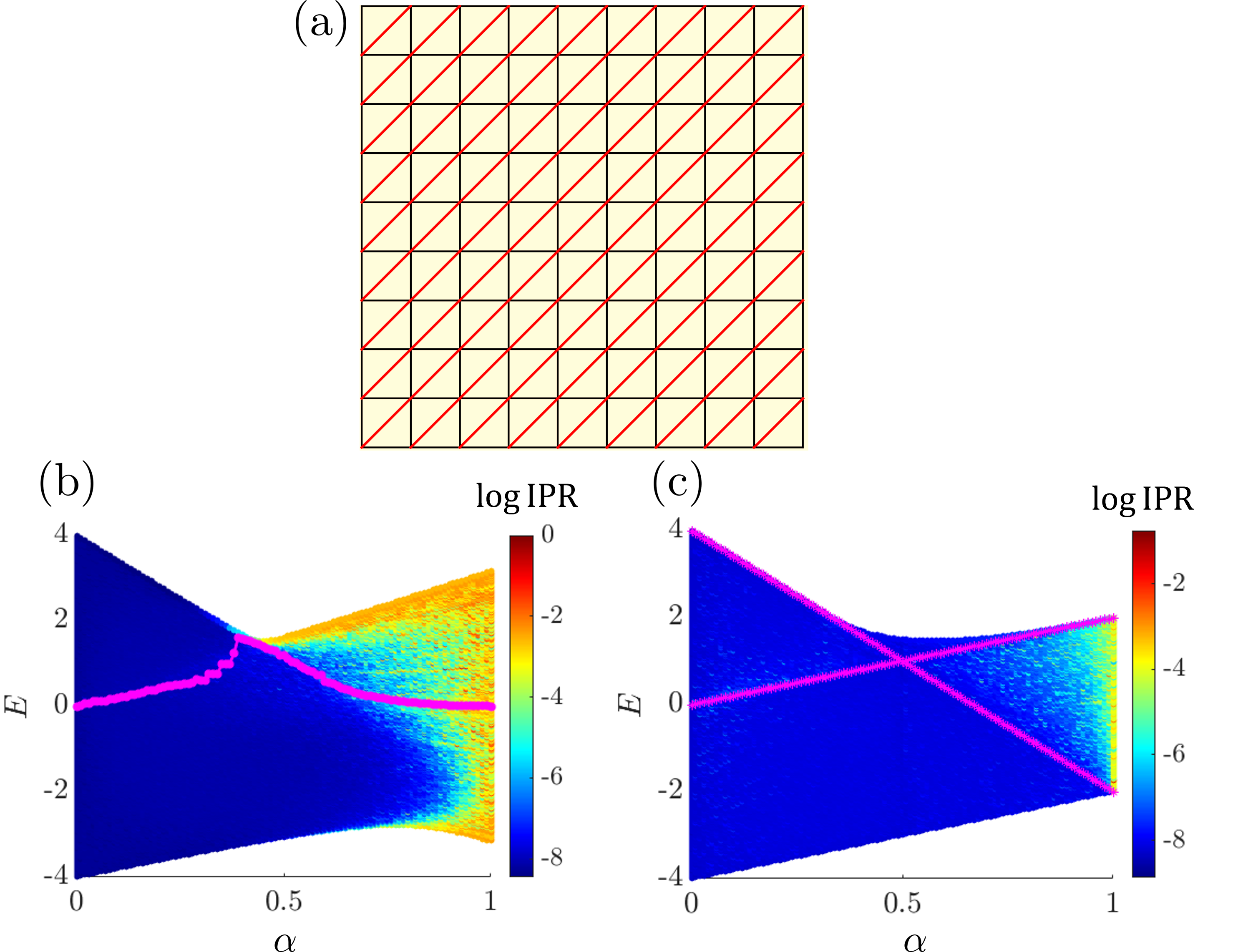}
    \caption{(a) Anisotropic triangular lattice as the periodic Truchet tiled system. The black and red links represent different hopping amplitudes $(1-\alpha)t$ and $\alpha t$ with $t=1$, respectively. The landscape of the IPR of the eigenstates in the (b) random Truchet tiling and (c) anisotropic triangular lattice. The magenta curves are drawn 
    %for emphasizing 
    to emphasize
    the energy at which the Van Hove singularities appear.}
    \label{fig: sup_AT}
\end{figure}

\section{Numerous zero modes in the strong disorder regime}
    \label{sec:2}
    Here, we identify two distinct origins for the emergence of numerous zero-energy states in the random Truchet tiling, as compared to the AT lattice, in the limit where $\alpha$ approaches unity: (i) sublattice imbalance, and (ii) the presence of internal symmetry within a sublattice. Note that in this limit, both horizontal and vertical hopping terms become negligible. Thus, the Hamiltonian becomes \begin{align}
    \label{Hsup}&\mathcal{H}=\lim_{\alpha\to1}H=-t\sum_{\braket{\braket{i,j}}_{\mathrm{diag}}\sigma}c_{i\sigma}^\dagger c_{j\sigma}+\mathrm{H.c.}
    \end{align}
Notably, the Hamiltonian $\mathcal{H}$ has chiral symmetry regardless of the specific details of the Truchet tiling pattern. Specifically, $\Gamma\mathcal{H}=-\mathcal{H}\Gamma$, where $\Gamma c_{i\sigma}= (-1)^{x(i)/a}c_{i\sigma}\Gamma$ and $\Gamma c_{i\sigma}^\dagger= (-1)^{x(i)/a}c_{i\sigma}^\dagger\Gamma$. Here, $x(i)=na$ for some integer $n$. $a$ is the 
%unit length scale 
lattice constant
of the square lattice. This chiral symmetry arises because each diagonal link connects sites at positions $a(n,m)$ and $a(n\pm1,n\pm1)$ or $a(n\pm1,m\mp1)$, where $n$ and $m$ are integers.
    
Due to the chiral symmetry, the matrix representation of $\mathcal{H}$ with proper choice of basis becomes $\begin{pmatrix} 0 & T \\ T^{\dagger} &  0\end{pmatrix}$, where $T$ is the hopping matrix. The number of rows and columns of $T$ are $N_A$ and $N_B$, which correspond to the number of sites in each sublattice, respectively. Note that $T$ doesn't need to be a square matrix. Let us assume that $N_A\le N_B$ without loss of generality. Then, the zero-energy states of $\mathcal{H}$ stem from either (i) sublattice imbalance ($N_A<N_B$) or (ii) presence of symmetry within a sublattice ($\mathrm{rank}(T)<N_A$).

%To understand how such zero-energy states emerge as the consequence of the destructive interference effect, we note that the matrix representation of $\mathcal{H}$ with proper choice of basis becomes $\begin{pmatrix} 0 & T \\ T^{\dagger} &  0\end{pmatrix}$, where $T$ is the hopping matrix due to the chiral symmetry. Therefore, the number of zero-energy states is given by the nullity of $T$. Since $T$ is the adjacency matrix of the graph, its nullity i

\subsection{Sublattice imbalance}
    
     The presence of chiral symmetry guarantees that there is a zero-energy state originating from the sublattice imbalance, $N_A<N_B$, due to Lieb's theorem\cite{lieb1989two}. For instance, if the bipartite graph formed by diagonal links has an odd number of sites, a zero-energy state is guaranteed. Let us focus on the random Truchet tiling pattern, which becomes many independent bipartite graphs composed of diagonal links as $\alpha\to1$ (see the red links in Fig.~\ref{fig: sup_alpha} (a)). Due to the randomness, in the thermodynamic limit, we have a large number of disconnected components comprised of diagonal links and an odd number of sites (Fig.~\ref{fig: sup_alpha} (b)) or an imbalanced number of sublattices (Fig.~\ref{fig: sup_alpha} (c)) in the bulk. As a result, we have numerous zero-energy states as $\alpha\to1$ in the random Truchet tiling (see Fig.~\ref{fig: sup_alpha} (e)).
\begin{figure}[h]
    \centering
    \includegraphics[width=0.6\textwidth]{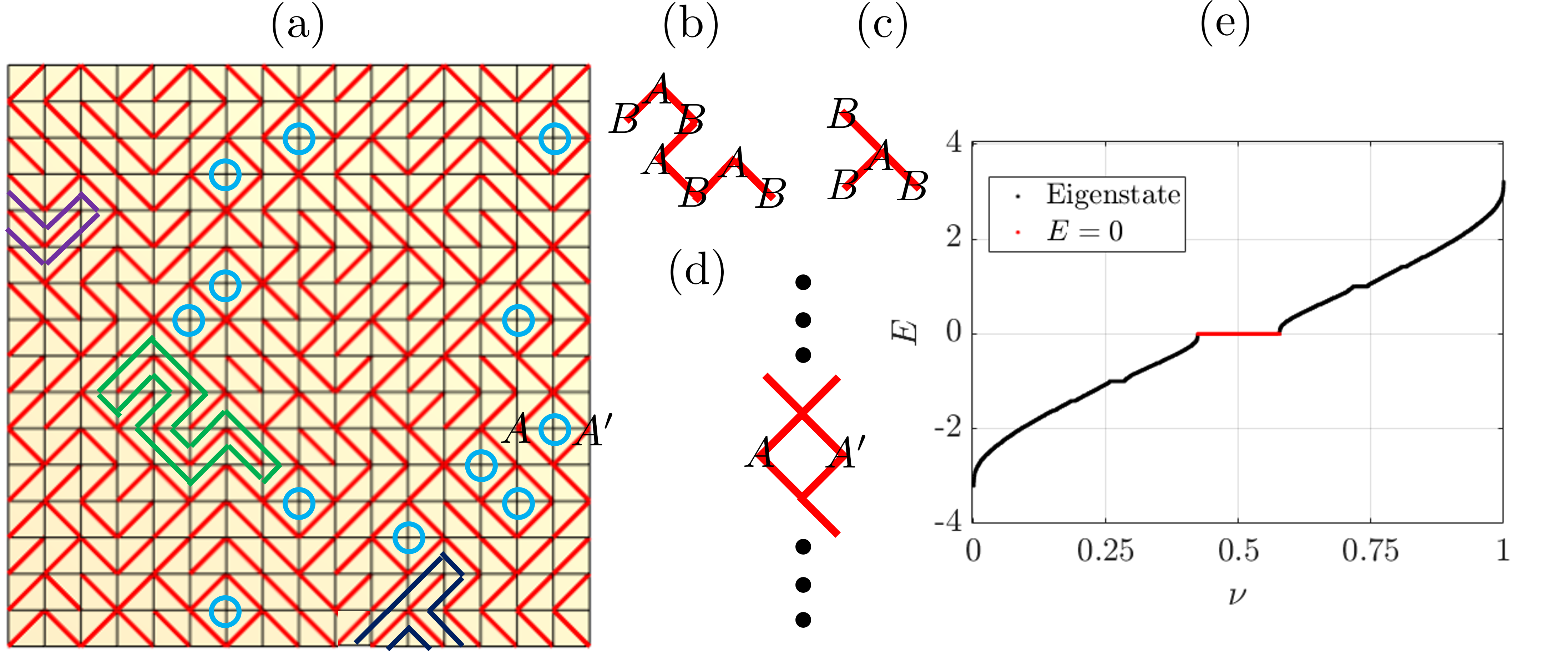}
    \caption{(a) Random Truchet tiling pattern. The skyblue, green, violet, and navy curves are drawn 
    to emphasize various graphs that have the zero-energy states due to the sublattice imbalance. $A$ and $A'$ are interchangeable vertices since they share the same adjacent matrix information. (b) Tree diagram where a zero-energy state appears due to the sublattice imbalance, where the $A$-sublattice has 3 sites, while the $B$-sublattice has 4 sites. (c) Tree diagram where two zero-energy states appear due to the sublattice imbalance, where the $A$-sublattice has 1 site, while the $B$-sublattice has 3 sites. (d) A zero-energy state emerges due to the destructive interference between the wavefunctions at $A$ and $A'$ sites. (e) Spectrum of the effective Hamiltonian $\mathcal{H}$ as a function of normalized integrated density of states, $\nu$. The red color is drawn 
    %for emphasizing 
    to emphasize
    the zero-energy state. The system size is $130\times 130$.}
    \label{fig: sup_alpha}
\end{figure}

    On the other hand, the AT lattice reduces to one-dimensional chains as $\alpha\to1$ (see the red links in Fig.~\ref{fig: sup_AT} (a)). Under the open boundary condition, the number of chains comprised of an odd number of sites is $L/2$ ($(L+1)/2$) when $L$ is even (odd). We emphasize that there is no zero-energy state other than these emergent in the AT lattice with $\alpha\approx1$.

\subsection{Symmetry within a sublattice}
    Unlike the AT lattice, we have another source of the zero-energy states in the random Truchet tiling---the destructive interference within a cycle. Note that the subgraphs composed of diagonal links in the random Truchet tiling would exhibit cycles, whereas all subgraphs are trees in the AT lattice. We emphasize that the nullity of $T$ would be larger than $\vert N_A-N_B\vert$ if $\mathrm{rank}(T)<N_A\le N_B$ due to the symmetry within a sublattice given by the automorphism. To be more specific, let us consider the local structure shown in Fig.~\ref{fig: sup_alpha} (d). Here, $A$ and $A'$ sites have two connectivities, respectively. Regardless of the detailed shape of the whole graph, the local square cycle geometry shown in Fig.~\ref{fig: sup_alpha} (d) leads to the zero-energy state due to the destructive interference. In detail, when two sites $A$ and $A'$ belonging to the same $A$ sublattice are assigned wavefunction amplitudes of $+1$ and $-1$, respectively, while all other sites are set to zero, destructive interference results in a zero-energy state localized within the square. Note that this kind of zero-energy states are widely studied as the Aharanov-Bohm cages under the flux insertion on the various tiling patterns\cite{vidal1998aharonov,kremer2020square,jeon2022length}. Since these zero-energy states stem from the presence of an automorphism that interchanges two sites $A$ and $A'$ within the same sublattice rather than sublattice imbalance, they have a different origin.

    We also note that the local square geometry shown in Fig.~\ref{fig: sup_alpha} (d) also guarantees the presence of a zero-energy localized state at the center site of the square (see skyblue circles in Fig.~\ref{fig: sup_alpha} (a)). This is because two orientations of the diagonal link are mutually exclusive in each plaquette.

\section{The influence of the boundary conditions}
In this section, we show that the influence of the boundary condition turns out to be negligible. Figure \ref{fig: rebuttal2} exhibits a spectrum comparing open (left panels) and periodic (right panels) boundary conditions for different $\alpha$ values. There is no significant difference in both the spectrum and inverse participation ratio (IPR) between the two boundary conditions. 
\begin{figure}[h]
    \centering
    \includegraphics[width=0.9\textwidth]{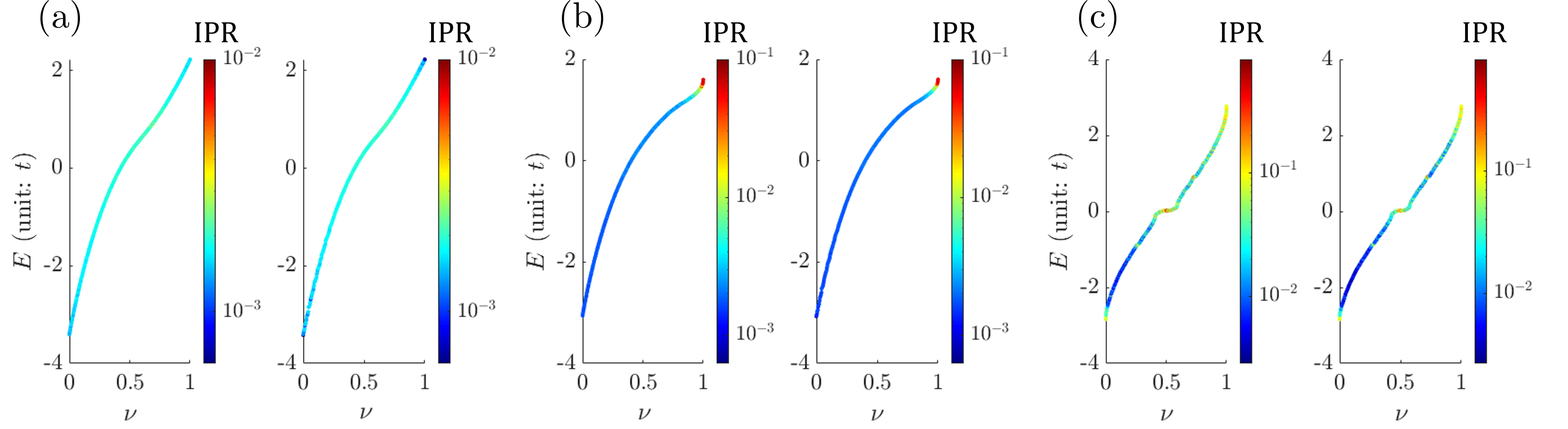}
    \caption{Comparison of the energy spectrum between the open (left panel) and periodic (right panel) boundary conditions for different $\alpha$ values: (a) 0.3, (b) 0.5, and (c) 0.9, respectively. The system size is $40\times 40$. 
    \label{fig: rebuttal2}}
\end{figure}

\section{Anderson localization of the random Truchet tiling}
Figure~\ref{fig: rebuttal4} illustrates the thermodynamic behavior of the spectrum for the localized regime. The normalized integrated density of states exhibits a smooth curve without stairs or plateaus [see Fig.~\ref{fig: rebuttal4} (a)]. This shows that the localization of the random Truchet tiling tends to be an Anderson insulator in the thermodynamic limit. Additionally, we investigate the fractal structure of the density of states by using the box counting dimension, $D_\mathrm{box}=\lim_{\delta\varepsilon\to0}\frac{\log(\mathcal{N}(\delta\varepsilon))}{\log(1/\delta\varepsilon)}$. Here, $\delta\varepsilon$ is the size of energy window, while $\mathcal{N}(\delta\varepsilon)$ is the number of occupied bins.  Figure~\ref{fig: rebuttal4} (b) demonstrates that the spectrum does not exhibit nontrivial fractal structure.
\begin{figure}[h]
    \centering
    \includegraphics[width=0.9\textwidth]{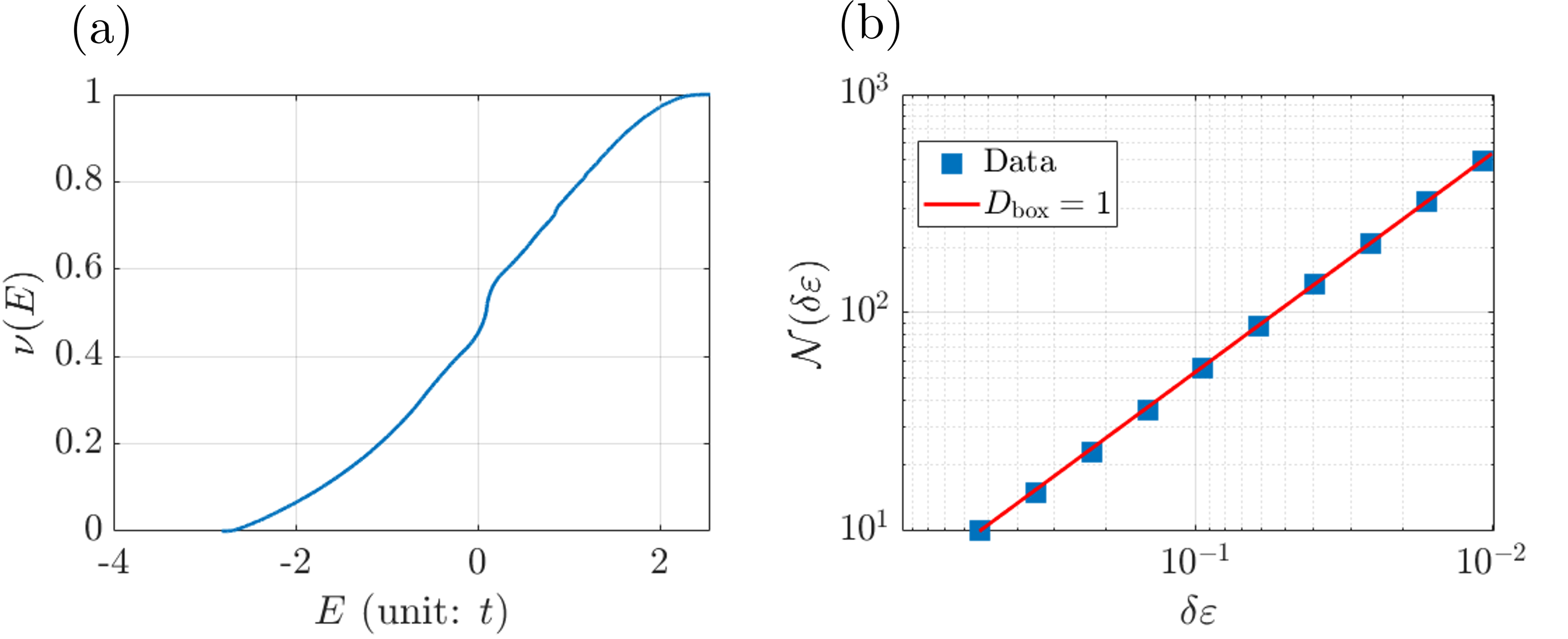}
    \caption{(a) Normalized integrated density of states, $\nu$ as the function of energy $E$ in the random Truchet tiling. Here, $\alpha=0.8$. The system size is $130\times 130$. The fractal structure or plateau is not observed. (b) The box counting dimension of the density of states. The unity box counting dimension $D_\mathrm{box}=1$ graph is drawn to emphasize that the spectrum is not fractal.
    \label{fig: rebuttal4}}
\end{figure}
\end{widetext}

\end{document}